\documentclass[%
 reprint,
nofootinbib,
 amsmath,amssymb,
 aps,
prd,
]{revtex4-2}

\usepackage{aas_macros}
\usepackage[T1]{fontenc}
\usepackage{amssymb}
\usepackage{xcolor}
\usepackage{graphicx, epsfig, bm, amsmath,mathtools}
\usepackage{dcolumn}
\usepackage{hyperref}
\usepackage{cleveref}
\usepackage{float}

\usepackage{caption}
\usepackage{subcaption}
\usepackage{soul}

\usepackage{url}

\begin{document}

\preprint{APS/123-QED}

\title{The Effect of Nonlinear Gravity on the Cosmological Background During Preheating}

\author{Ryn Grutkoski}
\affiliation{Department of Astronomy \& Astrophysics, The University of Chicago, Chicago, IL 60637, USA}

\author{Hayley J. Macpherson}
\affiliation{NASA Einstein Fellow}
\affiliation{Kavli Institute for Cosmological Physics, The University of Chicago, 5640 South Ellis Avenue, Chicago, Illinois 60637, USA}

\author{John T. Giblin, Jr.}
\affiliation{Department of Physics, Kenyon College, Gambier, Ohio 43022, USA}
\affiliation{CERCA/ISO, Department of Physics, Case Western Reserve University, 10900 Euclid Avenue, Cleveland, Ohio 44106, USA}

\author{Joshua Frieman}
\affiliation{Kavli Institute for Cosmological Physics, The University of Chicago, 5640 South Ellis Avenue, Chicago, Illinois 60637, USA}
\affiliation{Department of Astronomy \& Astrophysics, The University of Chicago, Chicago, IL 60637, USA}

\date{\today}

\begin{abstract}
We use numerical relativity to study the violent preheating era at the end of inflation. This epoch can result in highly nonlinear fluctuations in density and gravitational potential which feed back onto the averaged expansion rate---an effect known as backreaction. Usually, simulations of preheating use the Friedmann constraint to enforce the Hubble expansion of spacetime during the evolution. In numerical relativity, this is not required and the inhomogeneous spacetime is evolved self-consistently. 
For a `vanilla' preheating model, we find a violation of the Friedmann constraint at the level of 0.005\% over the entire simulation. This violation increases to $\sim10\%$ as we sample smaller scales in the simulation domain.
\end{abstract}

\maketitle


\section{\label{sec:intro}Introduction}
On large scales, the Universe is generally well-described by the $\Lambda$CDM framework---a spatially-flat, nearly homogeneous and isotropic model. Its contents are treated as consisting predominantly of cold dark matter (CDM) that can be approximated by a perfect, pressureless fluid and dark energy that behaves as a cosmological constant ($\Lambda$). The leading paradigm for generating the initial density field is cosmic inflation: an early epoch of near-exponential expansion~\cite{Guth:1980zm,Linde:1981mu,Albrecht:1982wi,Linde:1983gd}. When inflation ended, the energy density stored in the homogeneous mode of the inflaton---the field that drove inflation---transferred to relativistic degrees of freedom that subsequently thermalized. This process, known as reheating, created the conditions necessary for Big Bang Nucleosynthesis. However, the details of the inflationary epoch and the subsequent transition to a radiation-dominated, hot Big Bang are not completely understood.

While measurements of the cosmic microwave background (CMB) radiation and the large scale structure of the Universe have constrained possible inflationary models~\cite{Mueller:2021jvn,Cabass:2022wjy,Cabass:2022ymb,DAmico:2022gki,Philcox:2022frc}, there is still no broadly accepted model for inflation and reheating. In most models, the dynamics of the inflationary period are determined by the homogeneous value(s) of the field(s) in the model. In traditional models of reheating, the inflaton field undergoes perturbative decays to other particles as it oscillates about the minimum of its potential~\cite{Abbott:1982hn, Albrecht:1982mp, Allahverdi:2010xz}. 

However, a more violent behavior arises in many models as nonlinear processes take over---a phenomenon known as \textit{preheating}~\cite{Traschen:1990sw,Shtanov:1994ce,Kofman:1994rk, Kofman:1997yn, Allahverdi:2010xz}. In preheating, resonances amplify the production of particles and growth of inhomogeneities. This violent era results in large fluctuations in density and the gravitational potential, both of which can grow to order unity~\cite{Bassett:1998wg,Bassett:1999cg,Bassett:1999mt,Adshead:2023mvt,Giblin:2019nuv}. These amplified inhomogeneities lead to potentially observable relics such as gravitational waves~\cite{Khlebnikov:1997di,Easther:2006gt,Easther:2006vd,Garcia-Bellido:2007nns,Easther:2007vj,Dufaux:2007pt,Dufaux:2008dn,Dufaux:2010cf}, including stochastic gravitational wave backgrounds~\cite{Khlebnikov:1997di, Easther:2006gt, Easther:2006vd, Garcia-Bellido:2007nns, Easther:2007vj, Dufaux:2007pt, Dufaux:2008dn, Dufaux:2010cf, Adshead:2018doq, Adshead:2019lbr, Adshead:2019igv, Cosme:2022htl}, non-Gaussianity in the CMB~\cite{Enqvist:2004ey,Enqvist:2005qu,Enqvist:2005pg,Kolb:2005ux,Jokinen:2005by,Barnaby:2006cq,Barnaby:2006km,Chambers:2007se,Bond:2009xx}, primordial magnetic fields and baryon asymmetry~\cite{Adshead:2016iae,Anber:2015yca, Adshead:2015jza, Kamada:2016eeb, Caldwell:2017chz, Adshead:2017znw, Domcke:2019mnd, Domcke:2022kfs}, compact objects~\cite{Bassett:1998wg, Green:2000he, Jedamzik:2010dq, Martin:2019nuw, Musoke:2019ima, Auclair:2020csm, Eggemeier:2021smj,Bringmann:2011ut, Aslanyan:2015hmi}, including collapsed structures such as primordial black holes~\cite{Adshead:2023mvt,Bassett:1998wg,Jedamzik:2010dq,Martin:2019nuw,Auclair:2020csm}, or compact mini halos~\cite{Aslanyan:2015hmi,Bringmann:2011ut}. Future observations of these relics offer a potential window into the reheating epoch which can further constrain inflation models.

General Relativity (GR) provides the most accurate description of gravity and is a central component of the $\Lambda$CDM model. Due to the nonlinearity of Einstein's equations, exact solutions of GR can only be found by enforcing some kind of space-time symmetry. 
Numerically solving the full set of ten coupled GR equations is complicated and computationally expensive---with the first stable simulation of a black hole merger performed only within the last two decades~\cite{Pretorius:2005gq,Pretorius:2007jn}. For this reason, approximations have been a cornerstone of modeling cosmological and astrophysical processes. For example, the cosmological principle leads to the approximation of the Universe on large scales as exactly homogeneous and isotropic. In this case, the space-time metric reduces to the Friedmann-Lema\^itre-Robertson-Walker (FLRW) form, and the GR equations reduce to the Friedmann equations with a single degree of freedom: the cosmic scale factor, $a(t)$.

While the Universe is highly inhomogeneous on small scales, the adoption of the FLRW metric is typically justified by measurements of statistical homogeneity above scales of $\sim100 h^{-1}$Mpc~\cite{Hogg:2004vw,Scrimgeour_2012} and by the observed near-isotropy of the CMB~\cite{Planck:2020:isotropy}. These observations are conventionally interpreted to indicate that the large-scale amplitude of stress-energy perturbations and even more so the amplitude of metric perturbations around the FLRW metric are small. As a consequence, studies of departures from an FLRW universe are traditionally done using cosmological perturbation theory (CPT). 
To linear order in the metric perturbation, the Einstein equations can be decomposed into zeroth order equations---where the $00$-component gives a first-order differential equation for $a(t)$, which we refer to as the {\sl Friedmann constraint} and the average of the $ii$-components gives a second-order differential equation for $a(t)$, which we refer to as the {\sl Friedmann equation}---in which the evolution of $a(t)$ is determined by the spatially averaged density field,  plus first-order equations that describe the evolution of the coupled metric and stress-energy perturbations.

However, when we consider {\it non}linear GR, the above picture does not hold. In this case, we must use some kind of \textit{spatial averaging} procedure (whether explicit or implicit) to determine the zeroth order behavior of the large scale Universe. How to correctly smooth over small-scale nonlinearities to arrive at a large-scale average picture of the Universe is an unsolved problem in cosmology, e.g.~\cite{Ellis:2011,Clarkson:2011}.

In the most general picture, the spatially averaged evolution of an inhomogeneous universe does not match the evolution of FLRW universe with the same mean density (see~\cite{Buchert:2011sx} for a review). There are additional terms in the expression for the spatially averaged `Hubble' expansion that arise from smoothing over small-scale nonlinearities~\cite{Buchert:1999er}; this is what we refer to as cosmological \textit{backreaction}. These additional terms have been studied for the late Universe using perturbation theory, e.g.~\cite{LiSchwarz2007,ClarksonUmeh2011,Baumann2012}, exact solutions, e.g.~\cite{Rasanen2006,Garcia-Bellido2008,Bolejko2017}, and numerical simulations, e.g.~\cite{Adamek2015,Bentivegna2016,Macpherson:2019}---with the amplitude of the effect the subject of much debate, e.g.~\cite{GreenWald2014,Buchert2015}.

 Linear CPT has been used to study the dynamics of preheating~\cite{Frolov:2008hy,Huang:2011gf}. However, since preheating is an extremely nonlinear process, a linear treatment may not capture all the relevant physics~\cite{Giblin:2019nuv,Adshead:2023mvt}, including backreaction effects, which have never been studied in this era. These terms can be captured naturally via numerical relativity: a fully nonlinear computational method to solve Einstein's equations. 
 
Indeed, previous
investigations of preheating using numerical relativity simulations have confirmed the breakdown of CPT for modes with wavelengths comparable to the horizon size at the end of inflation~\cite{Giblin:2019nuv,Adshead:2023mvt}. 

Motivated by earlier studies of preheating using numerical relativity, our main goal in this work is to quantify the amplitude and evolution of backreaction effects during preheating and examine the scale-dependence of these effects. 
To do so, we use GABERel~\cite{Giblin:2019nuv}, the fully relativistic version of GABE~\cite{Child:2013ria}, to numerically evolve the complete set of Einstein's field equations and the scalar field equations of motion through parametric resonance preheating. 
We determine the impact of backreaction during this time by explicitly calculating the degree to which the Friedmann constraint, i.e. the first Friedmann equation, is violated.

In our simulations, we find evidence for non-zero backreaction effects during preheating. We show that these effects grow with the degree of nonlinearity of preheating and that they vary with spatial scale, with the smallest scales showing the largest effects. 
An exploration of whether these effects leave an imprint on the observational relics produced during preheating is warranted given that previous studies assumed the Friedmann equation for the cosmic expansion (with the exception of~\cite{Adshead:2023mvt}), which we leave to future work.

The paper is organized as follows: in Section~\ref{subsec:model} we define the model and introduce the formalism of Numerical Relativity. In Section~\ref{sec:calculations} we detail our calculations for determining the amplitude of backreaction effects. Results and discussion of our numerical simulations are presented in Section~\ref{results}, and final remarks along with proposed directions for future work are given in Section~\ref{sec:conclusion}.

Throughout this work, we use natural units such that $c=\hbar=1$, and define the Planck mass as $m_{\text{pl}}=1/\sqrt{G}$. Greek indices represent space-time indices, with values 0...3. Latin indices represent spatial indices, with values 1...3. Repeated indices are summed using the Einstein summation convention. 

\section{\label{sec:sim}Model and Formalism}
We are interested in studying nonlinear gravitational effects in preheating, with specific focus on their impact on violation of the Friedmann constraint. Thus, it is necessary to capture the full nonlinearity of Einstein's equations using numerical relativity. We first describe the model of preheating we use in Section~\ref{subsec:model}, then describe our approach to numerical relativity in Section~\ref{subsec:nr}, and finally detail our choice of initial conditions in Section~\ref{subsec:init}.

\subsection{\label{subsec:model}Model of preheating}
We use a canonical model of preheating---sometimes called {\sl vanilla preheating}---with two scalar fields: the inflaton $\varphi$ and a coupled, massless matter field $\chi$ \cite{Traschen:1990sw}. This system is in general described by the action
\begin{equation}
    S=\int {\rm d}^4x\sqrt{-g}\left(-\frac{1}{2}\,\partial^\mu\varphi\,\partial_\mu\varphi -\frac{1}{2}\,\partial^\mu\chi\partial_\mu\chi -V(\varphi,\chi)\right),
\end{equation}
where $g$ is the determinant of $g_{\mu\nu}$,  $V(\varphi, \chi)$ specifies the particular model of inflation and preheating being studied, and $\partial_\mu\equiv\partial/\partial x^\mu$. For this work, we assume a quadratic inflaton potential and four-leg coupling between the two fields,
\begin{equation}
    V = \frac{1}{2}m^2\varphi^2+g^2\varphi^2\chi^2,
    \label{potential}
\end{equation}
where we set the coupling strength to $g^2=2.5\times10^{-7}$ and the mass of the inflaton field to $m=10^{-6}m_{\text{pl}}$. This model has been extensively studied in the literature~\cite{Traschen:1990sw} and has often been used as a benchmark preheating scenario, making it a useful toy model despite being disfavored by both recent CMB data~\cite{Planck:2018vyg} and the necessity of extensions needed to fully deplete the inflaton's energy~\cite{Dufaux:2006ee,Amin:2014eta}. 
Since the dynamics of this model are well studied, it is a good starting place for our initial investigation of the effects of backreaction.

\begin{figure}[h]

    \includegraphics[width=0.95\linewidth]{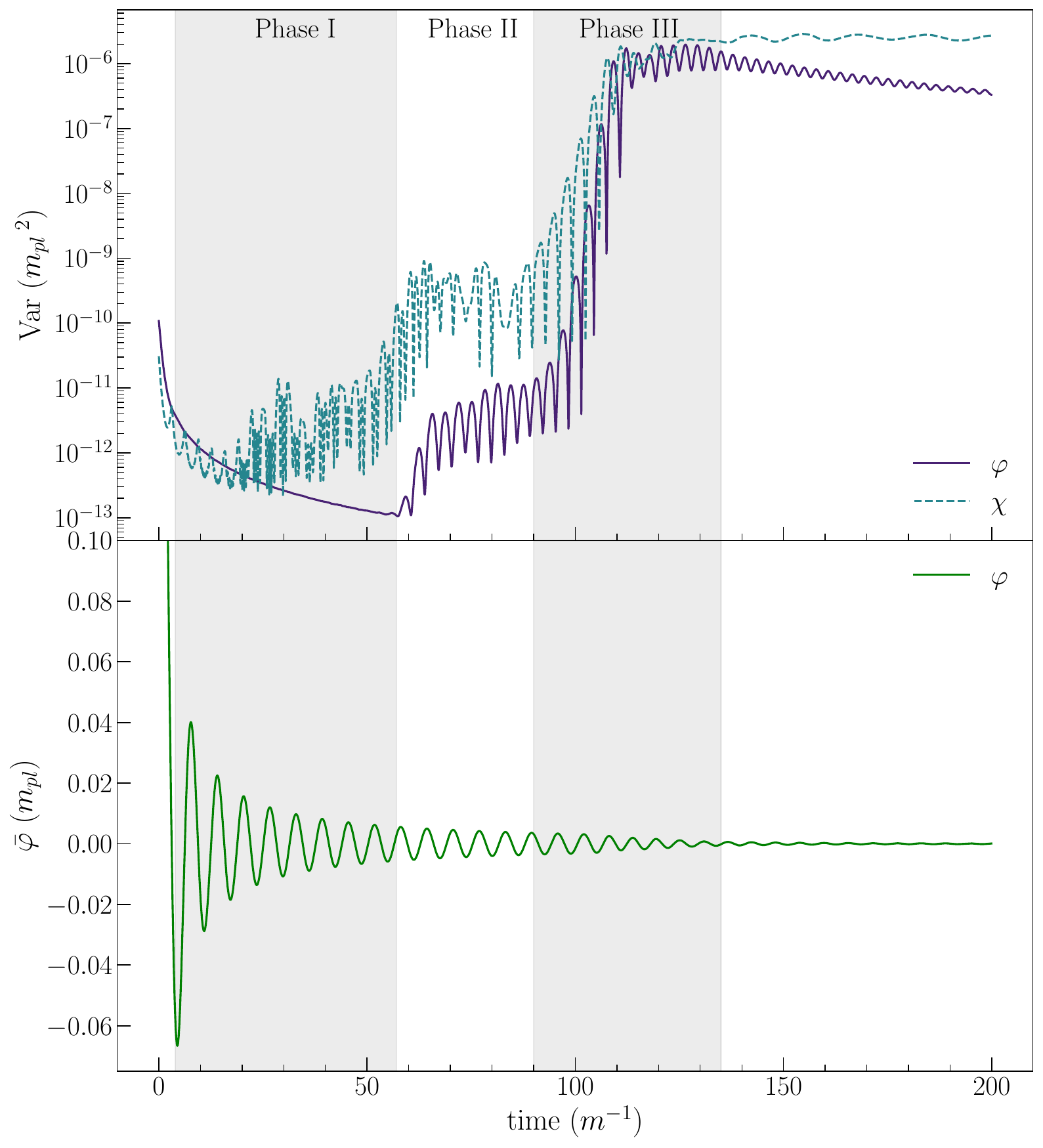}
    \caption{The stages of parametric resonance during preheating. The top panel shows evolution of the variances of the inflaton (purple, solid) and the coupled scalar field (teal, dashed) as a function of time. In Phase I, the first shaded region, the inflaton $\varphi$ variance oscillates, serving to amplify modes of the coupled scalar field $\chi$ through the quadratic coupling, but backscattering has yet to break down the inflaton condensate. In Phase II, the time between the two shaded regions, there are enough $\chi$ particles to backscatter into the inflaton field, which results in growth in the variances of both fields. In Phase III, the last shaded region, both field variances grow quickly as power is distributed over many modes. These three phases are characteristic of parametric resonance preheating. The bottom panel shows the average of the inflaton condensate as it oscillates around the bottom of its potential.}
    \label{fig:phases}
\end{figure}

In this model, when inflation ends the majority of the energy density of the Universe is contained in the inflaton condensate. The homogeneous mode of the inflaton field begins to oscillate around the minimum of its potential, as shown in the lower panel of Fig.~\ref{fig:phases}. This leads to parametric amplification of certain modes of the coupled matter field, $\chi$. This process continues until the produced $\chi$ particles scatter into the inflaton field, and nonlinear physics leads the condensate to break down.  The parametric resonance that occurs in this process can be broken into three distinct phases, which can be summarized as follows: 
\begin{itemize}
    \item {\bf Phase I.} The inflaton decays into $\chi$ particles, increasing the variance of the $\chi$ field.
    \item {\bf Phase II.} The condensate of $\chi$ particles becomes dense enough that the inflaton begins to scatter off it, increasing the fluctuations of both fields.
    \item {\bf Phase III.} Eventually the fields become sufficiently amplified that the process goes fully nonlinear, and power efficiently transfers between modes of both fields.
\end{itemize}
In the upper panel of Fig.~\ref{fig:phases} we show how these phases appear in the amplitudes of spatial fluctuations of the inflaton $\varphi$ (solid purple curve) and the coupled matter field $\chi$ (dashed teal curve) as a function of simulation time, $mt$; the phases described above are labeled by the shaded regions.
The details of the simulation we show in this figure can be found in Sections~\ref{subsec:nr} and~\ref{subsec:init}. The amplitude of spatial fluctuations, i.e. the variance, is defined as
\begin{equation}
    \label{eq:variance}
    \text{Var}(\varphi)=\langle\varphi^2\rangle-\langle\varphi\rangle^2,
\end{equation}
where $\varphi$ is the field value at each grid point and here, angled brackets represent a Euclidean spatial average\footnote{Strictly this average is defined as Eqn. \eqref{eq:avg_def} with $\gamma=1$. For more details on our spatial averaging procedure see Section~\ref{ssec:back}.}.

\subsection{\label{subsec:nr}Numerical Relativity}
We use the Baumgarte-Shapiro-Shibata-Nakamura (BSSN) formalism~\cite{Baumgarte:1998te,Shibata:1995we} of numerical relativity---an adaptation of the Arnowitt-Deser-Misner (ADM) formalism of GR~\cite{Arnowitt:1962hi}. 

BSSN adopts a 3+1 decomposition of space-time into a series of spatial hypersurfaces defined by their time-like normal vector $n^\mu$. Under this decomposition, the metric tensor is parameterized as
\begin{equation}
        g_{\mu\nu}=\begin{pmatrix}
        -\alpha^2+\beta_i\beta^i & \beta_j \\ \beta_i & \gamma_{ij}
        \end{pmatrix},
\end{equation}
where $\alpha$ and $\beta^i$ are the lapse function and shift vector, respectively, and are used to define the gauge degrees of freedom. 
The spatial metric of the three-dimensional spatial hypersurfaces is $\gamma_{ij}$, which is further decomposed into a conformal factor $\phi$ and unit-determinant spatial metric $\tilde{\gamma}_{ij}$, such that
\begin{equation}
    \gamma_{ij}=e^{4\phi}\tilde{\gamma}_{ij}.
\end{equation}

Alongside the evolution equations in the BSSN formalism, the 3+1 projection of Einstein's equations yields two constraint equations: the Hamiltonian and momentum constraints. These spatial equations must be satisfied at all times for the simulation to be a true solution to Einstein's equations and are typically used to set initial data.
The Hamiltonian constraint is
\begin{equation}\label{eq:ham}
  \mathcal{R} + K^2 - K_{ij}K^{ij} = \frac{16\pi\rho}{m_{\text{pl}}^2},
\end{equation}
where $K_{ij}$ is the extrinsic curvature of the spatial hypersurfaces, $K={\rm Tr}(K_{ij})$, $\mathcal{R}$ is the spatial Ricci scalar, and $\rho\equiv T_{\mu\nu}n^\mu n^\nu$ is the energy density projected into the spatial hypersurfaces. For a more thorough introduction to BSSN, including the evolution equations themselves, see~\cite{Baumgarte:2010ndz,Baumgarte:2021skc}.

\subsubsection{Gauge}\label{sssec:gauge}

Since the lapse $\alpha$ and the shift $\beta^i$ are both pure gauge degrees of freedom, we are free to choose them---including their evolution equations---according to conditions that are convenient for the particular problem of interest. For example, one might want a slicing in which an observer at fixed spatial coordinates moves along a geodesic, i.e., the synchronous comoving foliation: $\alpha=1$ and $\beta^i=0$ at all times. However, this is not useful for nonlinear cosmological simulations, as coordinates will cross when structures become nonlinear. In our case, we thus work with a slicing from the Bona-Mass\'o family of solutions, corresponding to a choice of evolution of the lapse function of the form
\begin{equation}
\label{eq:slicing-general}
    \partial_t\alpha=F(\alpha) K,
\end{equation}
where $F(\alpha)$ is an arbitrary function of the lapse only. In this work, we choose $F(\alpha)=-2\alpha$, which has the benefit of keeping the simulation time coordinate comparable to cosmic time such that the frequency of the homogeneous mode remains constant~\cite{Baumgarte:2021skc}.  
We also perform simulations with two additional slicing conditions, detailed in Appendix~\ref{appendix:slicing converge}, to ensure that our results are robust for different foliations.

The Bona-Mass\'o family of slicing conditions is well studied (see~\cite{Baumgarte:2021skc} for discussion). With a careful choice of $F(\alpha)$, this slicing condition can be constructed such that the lapse evolution reduces to FLRW, i.e. $\alpha(t) \rightarrow a(t)$ in the homogeneous limit~\cite{Adshead:2023mvt}, and thus the simulation time coordinate aligns with conformal time.

A non-zero shift is required when studying the formation of black holes in order to avoid singularities causing the code to crash. Previous studies of preheating have investigated the potential formation of primordial black holes among other gravitational relics~\cite{Adshead:2023mvt,Bassett:1998wg,Jedamzik:2010dq,Martin:2019nuw,Auclair:2020csm}. However, previous work~\cite{Giblin:2019nuv} using the same model of preheating and identical initial conditions to what we use here found no formation of primordial black holes (PBH). We thus do not expect any PBH formation in our simulations, so we set the shift to zero for simplicity of our post-processing analysis. 
Thus, our final fiducial slicing is
\begin{equation}
    \label{eq:main-slicing}
    \partial_t\alpha=2\alpha K.
\end{equation}
Note that even if black holes were to form, the choice of $\beta=0$ would simply cause the simulation to crash rather than give physically inconsistent results. 

\subsubsection{GABERel}\label{sssec:gabe}
We use GABERel, the relativistic version of GABE~\cite{Child:2013ria}, for our simulations. GABERel solves Einstein's equations in numerical relativity on a finite, expanding lattice under periodic
boundary conditions. The code is capable of handling both scalar and gauge fields~\cite{Adshead:2023mvt}, though here we only explore the effects of backreaction in the case of scalar fields. 

Our fiducial simulation setup uses a cubic grid of resolution $N=128$ and a comoving box size $L=5\,m^{-1}/e=1.84\times10^{6}\,m_\text{pl}^{-1}$, which we choose so that the box is subhorizon at the initial time but horizon-sized at the end of inflation. The time step is defined relative to the lattice spacing, $\Delta x=L/N$, such that $\Delta t=\Delta x/20$. Having a timestep that is a small fraction of our initial lattice spacing ensures both that we have good resolution and that we stay far from the regime in which causality becomes a significant consideration, i.e., we satisfy the Courant condition of $\Delta t < \Delta x$ for our specific problem. We also perform two lower-resolution simulations, with $N=64$ and $N=32$, to demonstrate numerical convergence (detailed in Appendix~\ref{appendix:res converge}), as well as a simulation using a larger box size of $L=10m^{-1}/e=3.68\times10^{6}m_\text{pl}^{-1}$ to ensure our results are robust to that choice (detailed in Appendix~\ref{appendix:box size}). 

\subsection{\label{subsec:init}Initial conditions} 
We want to both fully capture the effects of nonlinear gravity and also account for any effects that arise at the end of inflation, when the second time derivative of the scale factor crosses zero (i.e., $\ddot{a}=0$, where an over-dot is a derivative with respect to cosmic time). To do so, we start our simulation one $e$-fold before the end of inflation. For our chosen model, this corresponds to a homogeneous field value of 
$\bar{\varphi}\approx0.41 \, m_\text{pl}$~\cite{Giblin:2019nuv}. Importantly, starting before the end of inflation ensures that most of the simulation volume is subhorizon, while still letting us have horizon-scale modes at the end of inflation.
Beginning with a subhorizon volume ensures all modes begin with nearly Bunch-Davies initial conditions, given in Fourier space by
\begin{equation}
    \left\langle|\varphi(k)|^2\right\rangle_k=\frac{1}{2\omega},
\end{equation}
where $k$ is the wavenumber, $\omega=\sqrt{k^2+m_\text{eff}^2}$, $m_\text{eff}$ is the second derivative of the potential with respect to the field of interest, and $\langle\rangle_k$ is an average in Fourier space over all modes of magnitude $k$. 
We initialize both the $\varphi$ and $\chi$ fields in the same way: through the random generation of Gaussian-distributed power spectra for each component of each
field’s Fourier mode (see~\cite{Giblin:2019nuv} for a more detailed discussion). 

Since we are using a finite grid, there is a largest wavenumber in the simulation, known as the {\sl Nyquist mode} with $k_\text{Nyq}=\sqrt{3}\pi N/2L$. For modes near (or at) the Nyquist mode, finite-grid effect cause the simulation to diverge from the continuum system, numerical error will grow, and we cease to trust the simulations. While we cannot avoid this problem entirely, we can delay it by initializing the simulation with minimal power on these high-frequency scales--scales we have chosen to be irrelevant to the physics of the problem.
On the initial surface, we smoothly suppress power in modes with frequencies larger than a chosen cutoff frequency, $f_c$. This is controlled in our initial data by the parameter $\xi=f_c/f_\text{Nyq}<1$. We filter the power in all modes on the initial slice through a window function,
\begin{equation}
\label{eq:window}
    W(f)=\frac{1}{2}\left(1-\tanh\left[\kappa\left( f\frac{L}{2\pi}-\sqrt{3}N\xi\right)\right]\right).
\end{equation}
Here, $\kappa$ sets the sharpness of the cutoff; for this work we choose $\kappa=0.75$ and $\xi=1/8$. Throughout the simulation, we track the Hamiltonian constraint violation, which grows as power is transferred to small scales, see Appendix~\ref{appendix:constraint} for discussion of how we use this to ensure trust in the physical results of our simulation.

\section{\label{sec:calculations}Post processing}
In Section~\ref{ssec:back} we provide a brief background on the impact of backreaction on the spatially-averaged expansion, and in Section~\ref{ssec:subdom} we detail our procedure for calculating averages in spatial domains of varying size in our simulations.

\begin{figure*}
    \centering
    \includegraphics[width=\linewidth]{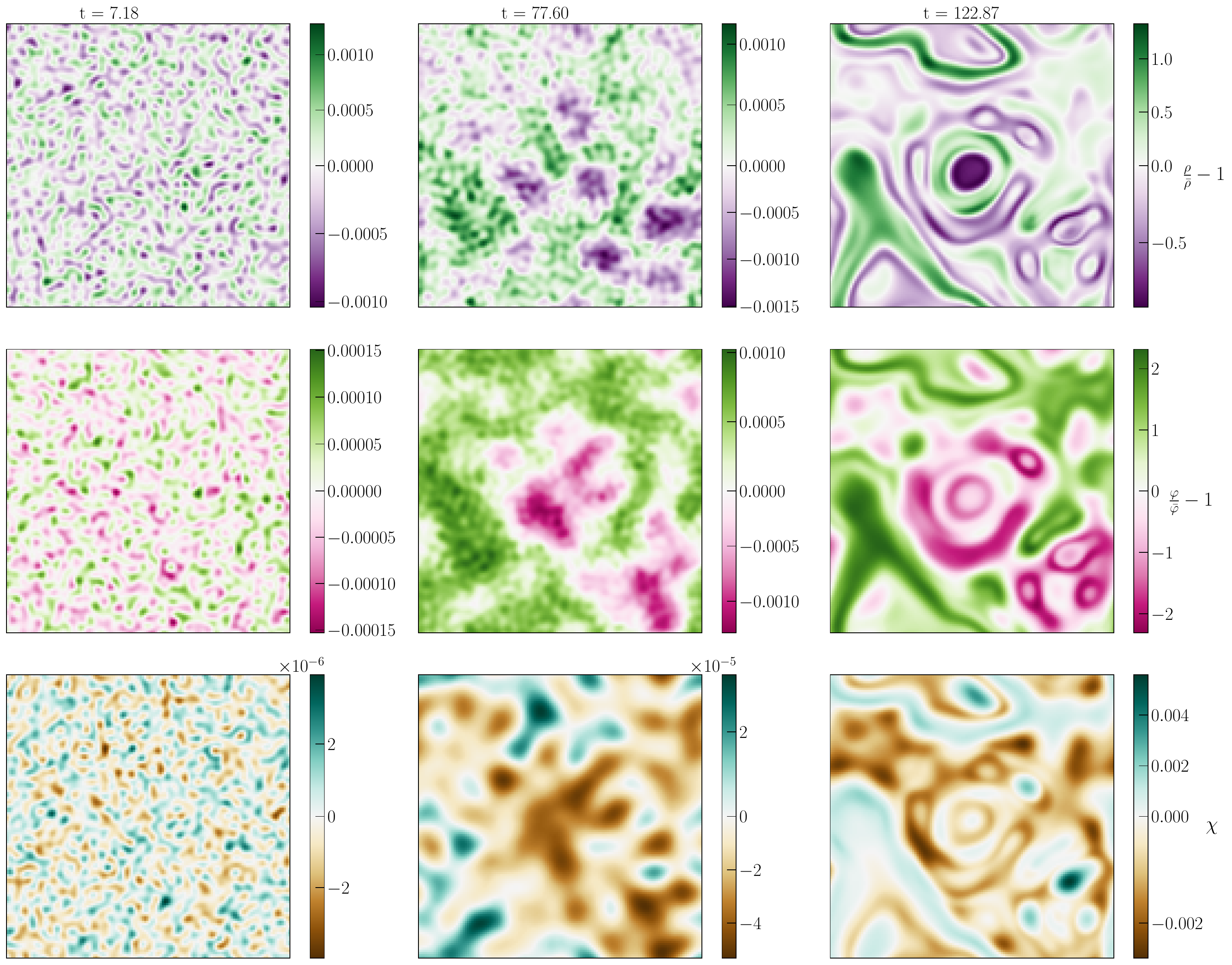}
    \caption{Two-dimensional ($128\times 128$) slices from the simulation volume: 
    the energy density, $\rho$ (top row), the inflaton amplitude, $\varphi$ (middle row), and the coupled scalar field amplitude, $\chi$ (bottom row) at times $t=7.18m^{-1}$ (left column), $t=77.6m^{-1}$ (middle column), and $t=122.9m^{-1}$ (right column). The quantities for $\rho$ and $\varphi$ are normalized by their average values at the given time.}
    \label{fig:2dplots}
\end{figure*}

\subsection{Averaging scheme}\label{ssec:back}
We are interested in assessing the validity of the Friedmann constraint in describing the average dynamics during the preheating era. In an FLRW space-time, the Hamiltonian constraint \eqref{eq:ham} reduces to the Friedmann equation
\begin{equation}\label{eq:Friedmann}
    {\mathcal{H}}^2 = \frac{8\pi \bar{\rho}}{3m_{\rm pl}^2},
\end{equation}
for a flat spatial geometry with $\mathcal{R}\propto k=0$, where the cosmic-time Hubble parameter is $\mathcal{H}=\dot{a}/a$. 
In simulations of preheating, typically the Hubble expansion is sourced via Eqn. \eqref{eq:Friedmann} using the density averaged over the whole box, $\bar{\rho}$. Here, the overbar denotes a Euclidean spatial average, i.e., Eqn. \eqref{eq:avg_def} with $\gamma=1$.

Now dropping the FLRW assumption, we can define the spatially-averaged effective Hubble expansion of the spatial hypersurfaces as (see, e.g.~\cite{Buchert:1999er})
\begin{equation}\label{eq:Hubdef}
    \mathcal{H}_{\mathcal{D}} \equiv \frac{1}{3}\langle K \rangle_\mathcal{D} = \frac{\partial_t a_\mathcal{D}}{a_\mathcal{D}},
\end{equation}
where we define the spatial average of a scalar quantity, $y$, over some arbitrary three-dimensional spatial domain, $\mathcal{D}$, as
\begin{equation}\label{eq:avg_def}
    \langle y \rangle_\mathcal{D} \equiv \frac{1}{V_\mathcal{D}}\int_\mathcal{D} y \sqrt{\gamma}\, d^3X.
\end{equation}
In the above, $\sqrt{\gamma}\,d^3X$ is the volume element of the constant-time hypersurfaces, $\gamma$ is the determinant of the spatial metric $\gamma_{ij}$, and the volume of the domain is defined as
\begin{equation}
    V_\mathcal{D} \equiv \int_\mathcal{D} \sqrt{\gamma}\, d^3X.
\end{equation}
In Eqn. \eqref{eq:Hubdef}, we have also defined an effective scale factor, $a_\mathcal{D}$---replacing the FLRW scale factor---defined as
\begin{equation}
    a_{\mathcal{D}}(t) \equiv \left(\frac{V_\mathcal{D}(t)}{V_\mathcal{D}(t_i)}\right)^{1/3}.
\end{equation}

The spatial average of the Hamiltonian constraint \eqref{eq:ham} can then be written as (see~\cite{Gasperini:2010})\footnote{This scheme is an extension to the averaging scheme of \citet{Buchert:1999er} for an arbitrary coordinate system and averaging domain that is advected in time along the hypersurface normal, $n^\mu$, rather than the fluid flow lines.}
\begin{equation}\label{eq:avgHam} 
    \mathcal{H}^2_\mathcal{D} = \frac{8\pi\langle\rho\rangle_\mathcal{D}}{3m_{\rm pl}^2} - \frac{1}{6}\bigg(\langle\mathcal{R}\rangle_\mathcal{D} + \mathcal{Q_D}\bigg),
\end{equation}
where the \textit{kinematical backreaction} term $\mathcal{Q_D}$ arises because of the non-commutativity of averaging and time evolution in nonlinear GR; it is defined as
\begin{equation}
    \mathcal{Q_D} \equiv \frac{2}{3}\bigg( \langle K^2 \rangle_\mathcal{D} - \langle K \rangle_\mathcal{D}^2 \bigg) - 2 \langle A^2 \rangle_\mathcal{D},
\end{equation}
where $A^2 \equiv \frac{1}{2} A_{ij} A^{ij}$ and $A_{ij}$ is the trace-free part of the extrinsic curvature.

In this work, we are interested in a preliminary investigation of deviations from Friedmann evolution, rather than a detailed study of the amplitudes of the curvature and kinematical backreaction contributions. 
Thus, we first re-write Eqn. \eqref{eq:avgHam} as
\begin{equation}
    1 = \frac{8\pi\langle\rho\rangle_\mathcal{D}}{3 m_{\rm pl}^2\mathcal{H}^2_\mathcal{D}} - \frac{1}{6 \mathcal{H}^2_\mathcal{D}}\bigg(\langle\mathcal{R}\rangle_\mathcal{D} + \mathcal{Q_D}\bigg),
\end{equation}
as is typically done in defining standard cosmological parameters. 
From this, and using Eqn.~\eqref{eq:Hubdef}, we define
\begin{equation}\label{eq:Bdef}
    \mathcal{B}\equiv\frac{24\pi\langle\rho\rangle_\mathcal{D}}{\langle K \rangle_\mathcal{D}^2 m_{\rm pl}^2} - 1,
\end{equation}
which contains both $\langle\mathcal{R}\rangle_\mathcal{D}$ and $\mathcal{Q_D}$ and vanishes in the FLRW limit (or that of linear perturbations).
We calculate $\mathcal{B}$
on every time slice of the simulation to track how it evolves through the entire period of preheating.

It is important to note that any spatial averaging procedure explicitly depends on the chosen foliation of space-time into spatial sections, i.e. the gauge chosen for our simulations. In Appendix~\ref{appendix:slicing converge} we show that our results for $\mathcal{B}$ are robust to different, yet similar, choices in foliation. For a vastly different slicing of space-time we might expect our results to change. However, here we are interested in the level of backreaction in simulations of preheating, which typically adopt gauges similar to the ones we choose.

\subsection{Subdomain sampling procedure}\label{ssec:subdom}
The calculations for both $\langle K \rangle_\mathcal{D}$ and $\langle\rho\rangle_\mathcal{D}$ in Eq~\ref{eq:Bdef} can in principle be performed over any arbitrary spatial domain $\mathcal{D}$. We choose to perform the spatial average over \textit{both} the full simulation volume and smaller cubic subdomains in order to assess the level of backreaction on different physical scales (as was done in~\cite{Macpherson:2019} for a late Universe study of backreaction). 

We randomly select a number of subdomains from the full volume in order to beat down statistical noise. 
If a chosen subdomain crosses the boundary of the simulation volume, we employ the simulations periodic boundary condition and allow the subdomain to wrap. 
Since subdomains can be centered on any of the simulation grid points, the maximum number of non-identical subdomains of any given size is $N^3$, where $N$ is the resolution (number of grid points per box edge). As the subdomains are placed at random, they can overlap with each other, so they are not independent. Therefore, as the number of randomly selected subdomains grows very large, we are not adding new information; we therefore select 5\% of the possible subdomains of each size to sample and stipulate that no two of them overlap completely. For $N=128$, this corresponds to 104,857 subdomains of each size: 8, 16, 32, 64, and 96. 

\begin{figure}
    \centering    \includegraphics[width=0.95\linewidth]{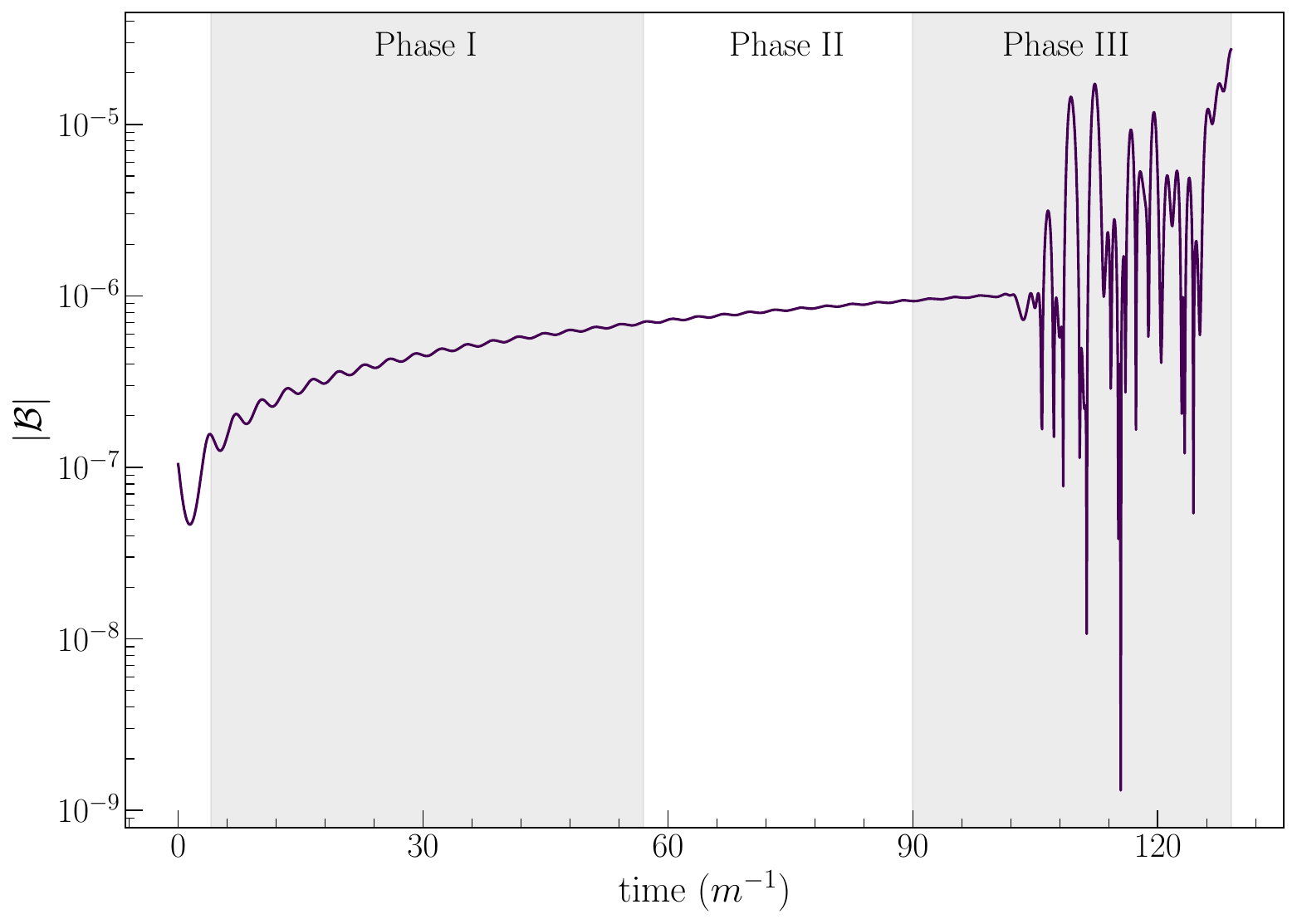}
    \caption{The value of the backreaction measure, $\mathcal{B}$, for the full simulation volume vs. time, also showing the three phases of preheating discussed in Section~\ref{subsec:model}. At late times, during the highly nonlinear phase of preheating, the behavior of $\mathcal{B}$ changes qualitatively, as it oscillates and grows rapidly.}
    \label{fig:full-vol}
\end{figure}

We calculate the median value of $\mathcal{B}$ and the probability distribution of $\mathcal{B}$ values for subdomains of varying size at each time step of the simulation. We use these statistical measures of $\mathcal{B}$ as a function of time to assess the violation of the Friedmann constraint on different spatial scales in the simulation.
We used the bootstrap method to calculate the uncertainty on our median values of $\mathcal{B}$. This method draws `new' batches of subdomains from the original subdomain sample (i.e., with some repeated values) and calculates the median for each new batch. We use the \texttt{SciPy} \textit{bootstrap} function with the percentile method and resample size of 500. This gives us an estimate of how much our median results would change for a different sample of the same number of subdomains, thus ensuring our chosen number of subdomains is sufficient.

The subdomains we choose are placed at random coordinate positions on the grid that do not change in time. The frame of the averaging procedure described in Section~\ref{ssec:back} is thus the simulation frame, \textit{not} the frame of the effective `fluid'. This means that mass is not conserved within individual subdomain volumes we consider, which might raise the question of the physical validity of these spatial domains and their explicit dependence on the (arbitrary) gauge choices we make for our simulations (see Section~3.5.5 of~\cite{Buchert:2020} for a discussion on this). 
However, we are interested in studying \textit{statistical} measures of the distribution of subdomain averages as a function of time; not the evolution of any one individual subdomain. 
The calculation at each time step should thus be considered as an independent `snapshot' of the statistics of the violation throughout the box rather than the evolution of a collection of subdomains.

\begin{figure*}
    \centering    \includegraphics[width=0.9\linewidth]{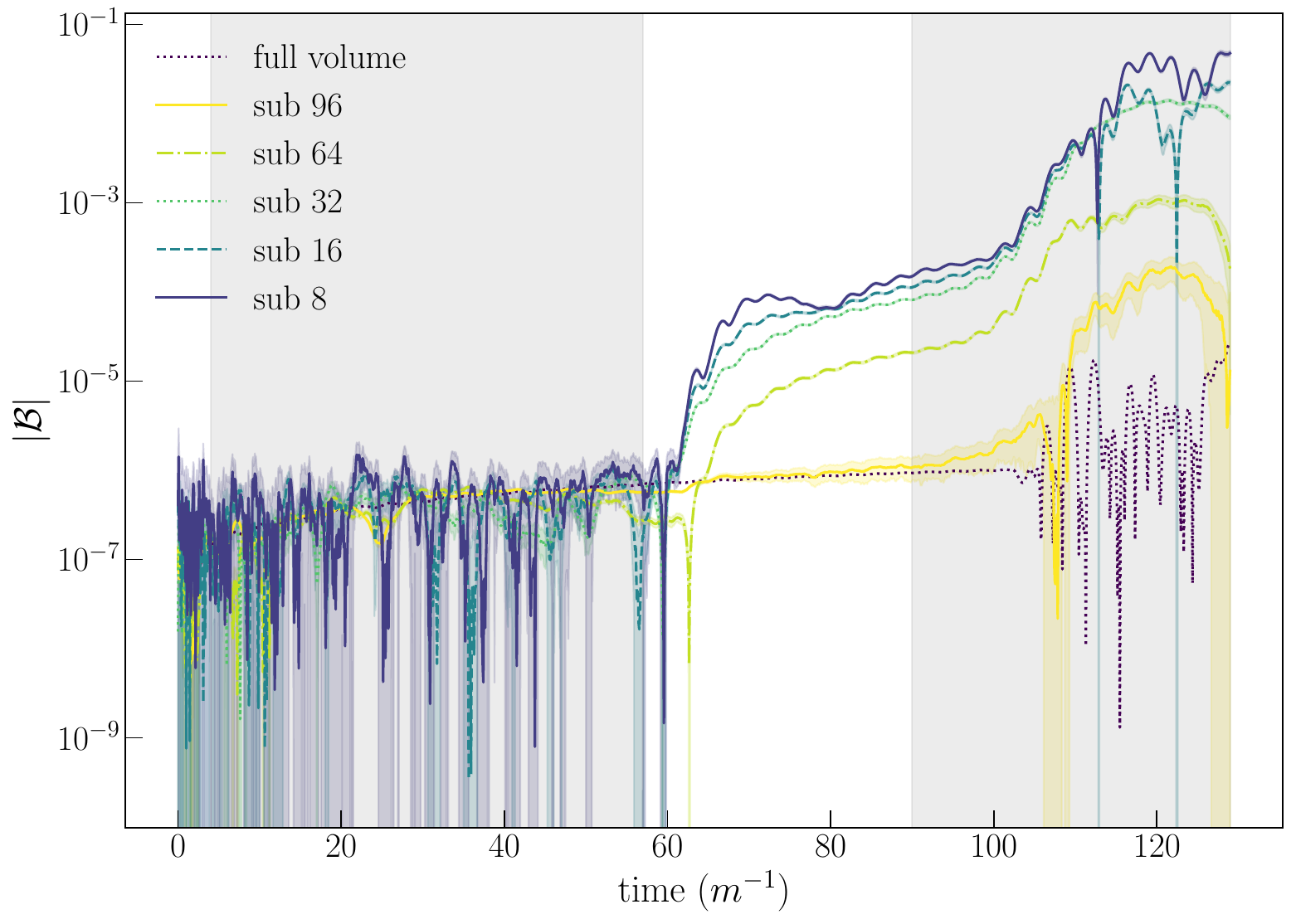}
    \caption{Evolution of the median value of the backreaction measure $\mathcal{B}$  for different subdomain sizes, along with the value of $\mathcal{B}$ for the full simulation volume. The gray shaded regions denote the phases discussed in~\ref{subsec:model}. The subdomain volumes are  $8^3$ cells (solid dark indigo line), $16^3$ (dashed blue line), $32^3$ (dotted teal line), $64^3$ (dot-dashed light green line), $96^3$ (solid yellow line), and the full $128^3$ volume (dotted purple line). Shaded regions colored corresponding to the curves denote the 95\% confidence intervals, calculated using the bootstrap method.}
    \label{fig:subs}
\end{figure*}

\begin{figure*}
    \centering
    \includegraphics[width=1\linewidth]{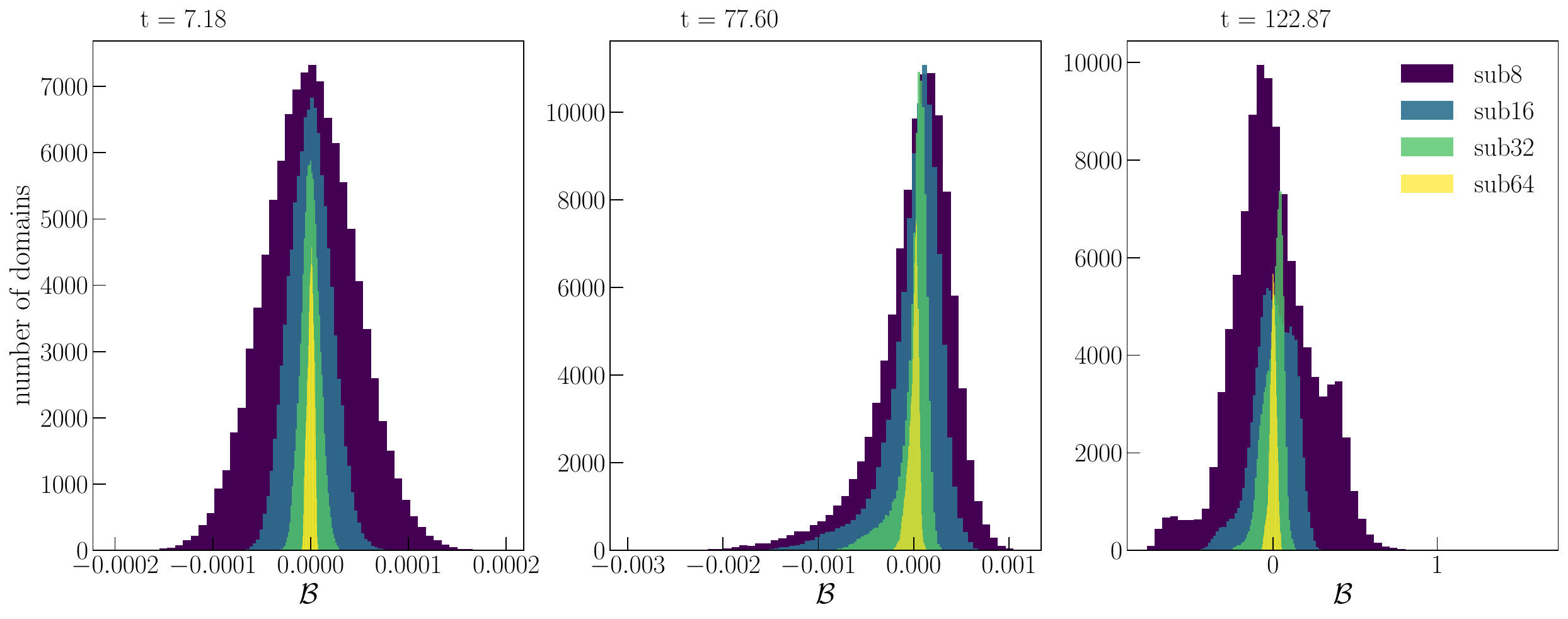}
    \caption{These panels show the distribution of $\mathcal{B}$ for different sized subdomains at three different times in the simulation. In each panel, the distribution of $\mathcal{B}$ for the $8^3$ subdomain is shown in purple, the distribution for $16^3$ in blue, for $32^3$ in green, and for $64^3$ in yellow. The left panel shows the distributions at an early time in Phase I, $t=7.18$, while the center panel shows them at time during Phase II, $t=77.6$, and the right panel shows them at a time in Phase III near the end of preheating, $t=122.87$. In the center and right panels, the skew of the distributions for $\mathcal{B}$ on the smaller subdomains is evident.}
    \label{fig:histogram}
\end{figure*}

\section{Results and discussion}\label{results}
In Fig.~\ref{fig:2dplots} we show two-dimensional slices of the full volume demonstrating the state of the simulation at three snapshots in time (left-to-right: $t=7.18m^{-1},$ $t=77.6m^{-1},$ and $t=122.9m^{-1}$). Top panels show the density, $\rho$, the middle panels show the inflaton field amplitude, $\varphi$, and the bottom panels show the coupled scalar field amplitude, $\chi$. The three time snapshots correspond respectively to the start of Phase I, the middle of Phase II, and near the end of Phase III. Moving from left to right panels, we see that the spatial fluctuations grow with time in both scale and amplitude for all three quantities. In the rightmost panels, the fluctuations have become non-linear, demonstrated by their large amplitude as indicated in each color bar. The scale of spatial fluctuation depends on the modes amplified by the parametric resonance in the earlier phases of preheating.

In the following sections, we will discuss our results calculating spatial averages in the simulation shown in Fig.~\ref{fig:2dplots}.

\subsection{Full volume averaging}
Fig.~\ref{fig:full-vol} shows the time evolution of $\mathcal{B}$ calculated over the full simulation volume, with the three phases of preheating discussed in Section~\ref{subsec:model} shown with shaded regions. We expect to see the backreaction effect increase in the later part of preheating, when the fluctuations become nonlinear (see Fig.~\ref{fig:2dplots}). 
Fig.~\ref{fig:full-vol} indeed shows that the backreaction effect increases in the final phase of preheating in this model. In the first two phases, $\mathcal{B}$ grows slowly and monotonically from $\sim 10^{-7}$ to $\sim 10^{-6}$; in the third phase, it oscillates dramatically and then grows rapidly to an amplitude of $\sim 10^{-5}$. 

Note that for this plot and all following plots we cut our analysis off at $t=130$. While the simulation remains stable beyond this time, numerical error causes constraint violation to grow unsatisfactorily beyond this time (see Appendix~\ref{appendix:constraint} for further discussion).

We have confirmed that these results are physical and not a numerical artifact due to resolution (Appendix~\ref{appendix:res converge}) or constraint violation from power transferring to small scales (Appendix~\ref{appendix:constraint}). 
We find that the backreaction is comparable for subdomains of the same physical scale, regardless of box size, as detailed in Appendix~\ref{appendix:box size}. We check two additional foliations, finding that the backreaction effect does not change significantly for different conditions, as detailed in Appendix~\ref{appendix:slicing converge}.

\subsection{Subdomain averaging}
In Fig.~\ref{fig:subs} we show the evolution of the median value of $\mathcal{B}$ for five different subdomain sizes, with side lengths of $n=$ 8, 16, 32, 64, and 96 grid cells, along with $\mathcal{B}$ for the full simulation volume. The colored shaded regions around each curve denote the 95\% confidence intervals, calculated using the bootstrap method. 

During Phase I, all subdomains show a very similar amplitude of $\mathcal{B}$ with fluctuations in time not seen in the full volume. These fluctuations tend to increase with decreasing size of the subdomain volume, with volumes of $8,16,$ and 32 grid cells having similar amplitude fluctuations. This can most likely be explained by the characteristic resonance scale during Phase I: when averaging on scales below $\sim32$ grid cells we are below the resonant wavelength of this phase, thus we expect high amplitude fluctuations in the average over subdomains of this size. In contrast, above this scale (for volumes of 64 and 96 grid cells) we see fewer fluctuations in time since when we average \textit{above} the resonant scale the oscillations are damped.

During Phase II, the amplitude of $\mathcal{B}$ grows by one or two orders of magnitude on small scales, but remains nearly constant for the largest subdomain (96 grid cells) and for the full volume. During Phase III, the amplitude grows further for all subdomains, with larger final amplitude for smaller subdomains: from smallest to largest subdomain, the final backreaction amplitudes are approximately $\mathcal{B}\approx0.05,\, 0.02, \,10^{-2},\, 10^{-4}$, and $10^{-5}$, where the latter is comparable to the value for the full simulation volume. 
Thus, we see a smooth reduction in $\mathcal{B}$ as we increase the size of the domain of averaging above 32 grid cells to the full volume.

Interestingly, the order of magnitude of $\mathcal{B}$ does not change for the three smallest subdomain sizes. This might indicate that, once the domain size is at or below $\sim32$ grid cells, we are averaging below scales of the most significant inhomogeneity in the simulation, i.e. the characteristic resonant scale (this can somewhat be seen by eye in the rightmost panels of Fig.~\ref{fig:2dplots}). This aligns with the explanation for the fluctuations in time we see during Phase I, which are most pronounced for these same domain sizes.

In Fig.~\ref{fig:histogram}, we show the distribution of $\mathcal{B}$ across different subdomain sizes at three different times, corresponding to the Phases I, II, and III described in Section~\ref{subsec:model}. The leftmost panel shows the distribution at time $t=7.18$, the center panel at $t=77.6$, and the rightmost panel at $122.87$. Each panel shows the distribution of $\mathcal{B}$ on the smallest subdomains, $8^3$, in purple, the next smallest $16^3$ in blue, the quarter-volume $32^3$ in green, and the half-volume $64^3$ in yellow. In the center and right panels, the increasing skew of the distribution is evident in the smaller subdomains, as we might expect as the scale of averaging approaches the nonlinear scale. Due to this skew at later times, the mean is no longer a good measure of the typical value, and so we have shown the median of $\mathcal{B}$ for all subdomain plots.

The backreaction measure $\mathcal{B}$ contains contributions from both the average scalar curvature $\langle\mathcal{R}\rangle_\mathcal{D}$ and the kinematical backreaction $\mathcal{Q_D}$. The former comprises second spatial derivatives of the metric and is thus expected to roughly follow the order of magnitude of density contrasts (through the lens of CPT), while the latter is related to variance in the extrinsic curvature---i.e. first time derivative of the metric. As the simulation becomes nonlinear, we thus expect the curvature term to be the dominant contribution to $\mathcal{B}$. As expected based on this logic, we can see the magnitude of $\mathcal{B}$ in Fig.~\ref{fig:histogram} to roughly follow that of density contrasts in the top row of Fig.~\ref{fig:2dplots} (where the three columns represent the same time snapshots as Fig.~\ref{fig:histogram}).

\section{\label{sec:conclusion}Conclusions}
Here we have presented the first investigation of the validity of a widely-used assumption in simulations of preheating: the Friedmann constraint. Using simulations of `vanilla' preheating, we found violation of the Friedmann constraint of order $5\times10^{-5}$ when averaging across the full simulation domain. We have shown that on small subdomains, this violation reaches order 10\% by the end of preheating. 

We have thus shown that not all the relevant physics is being captured by simulations that use FLRW and CPT, which enforce the Friedmann constraint. Even on the full simulation volume, the violation found is above the simulation error, indicating a need for further investigation.

The physics of preheating can change the predicted gravitational wave background after inflation, primordial non-Gaussianity signatures, and potentially the existence of primordial black holes. In general, large changes to the expansion history during this period have the potential to alter these observational predictions. Here we have found the expansion history remains close to FLRW, within $0.005\%$ across the full simulation volume; thus we expect any resulting changes in the observational consequences of preheating to also be very small.
We found violations of the Friedmann constraint at the $\sim$5--10\% level for subdomains at and below the characteristic resonant scale of the simulation. Whether such a large violation on smaller scales could lead to changes in observational predictions from preheating is worth exploring.

Further study can expand on this to explore the violation of FLRW in different inflationary and preheating models. The model used here, defined by Eqn.~\ref{potential}, has merit in being well-studied and well-understood, but it is disfavored both by observation~\cite{Planck:2018vyg} and theory~\cite{Dufaux:2006ee}. More notably, it is not one of the models expected to have the largest metric perturbations---models that have stronger instabilities, such as tachyonic instabilities, near the horizon might be affected by backreaction more substantially~\cite{Bassett:1998wg,Bassett:1999cg,Finelli:2000ya,Felder:2000hj,Felder:2001kt,Copeland:2002ku,Barnaby:2006cq}.

This work has served as an important proof-of-concept that the Friedmann constraint \textit{is broken} in the presence of strong nonlinearity. While the level of violation we find here is small, this conclusion is not necessarily applicable across the broad range of inflation models. In particular, many other models have stronger nonlinearities and thus we might expect a stronger violation of FLRW. Further study in these models is important to ensure our numerical modeling of this era of the Universe is accurate.

\begin{acknowledgments}
We thank Austin Joyce for helpful discussions throughout this work. RG is supported in part by the National Science Foundation Graduate Research Fellowship, Grant Number 2140001. Support for HJM was provided by NASA through the NASA Hubble Fellowship grant HST-HF2-51514.001-A awarded by the Space Telescope Science Institute, which is operated by the Association of Universities for Research in Astronomy, Inc., for NASA, under contract NAS5-26555. J.T.G. is supported in part by the National Science Foundation, PHY-2309919. Simulations were performed on hardware provided by the National Science Foundation, Kenyon College, and the Kenyon College Department of Physics. 
\end{acknowledgments}

\appendix

\section{Numerical robustness tests}\label{appendix:robustness}
It is important to ensure that our results are physical and not dominated by numerical error. To do so, we look at three different numerical choices: the resolution, the box size, and the slicing condition.

\subsection{Resolution}\label{appendix:res converge}
We perform our primary simulation, detailed in Section~\ref{sec:sim}, for three different resolutions, namely $N=32$, $N=64$, and $N=128$. If our result is dominated by numerical error, we expect it to change as we change the resolution. If the result is physical, we expect it to be robust to changes in numerical resolution (so long as the physical problem remains largely unchanged). To hold the physical set-up unchanged, we adjust the window function, given by Eqn.~\eqref{eq:window}, setting $\xi=1/32$ for $N=128$, $\xi=1/16$ for $N=64$, and $\xi=1/8$ for $N=32$ such that the initial power spectrum cutoff is the same physical scale regardless of resolution.
 
\begin{figure}
    \centering
    \includegraphics[width=0.95\linewidth]{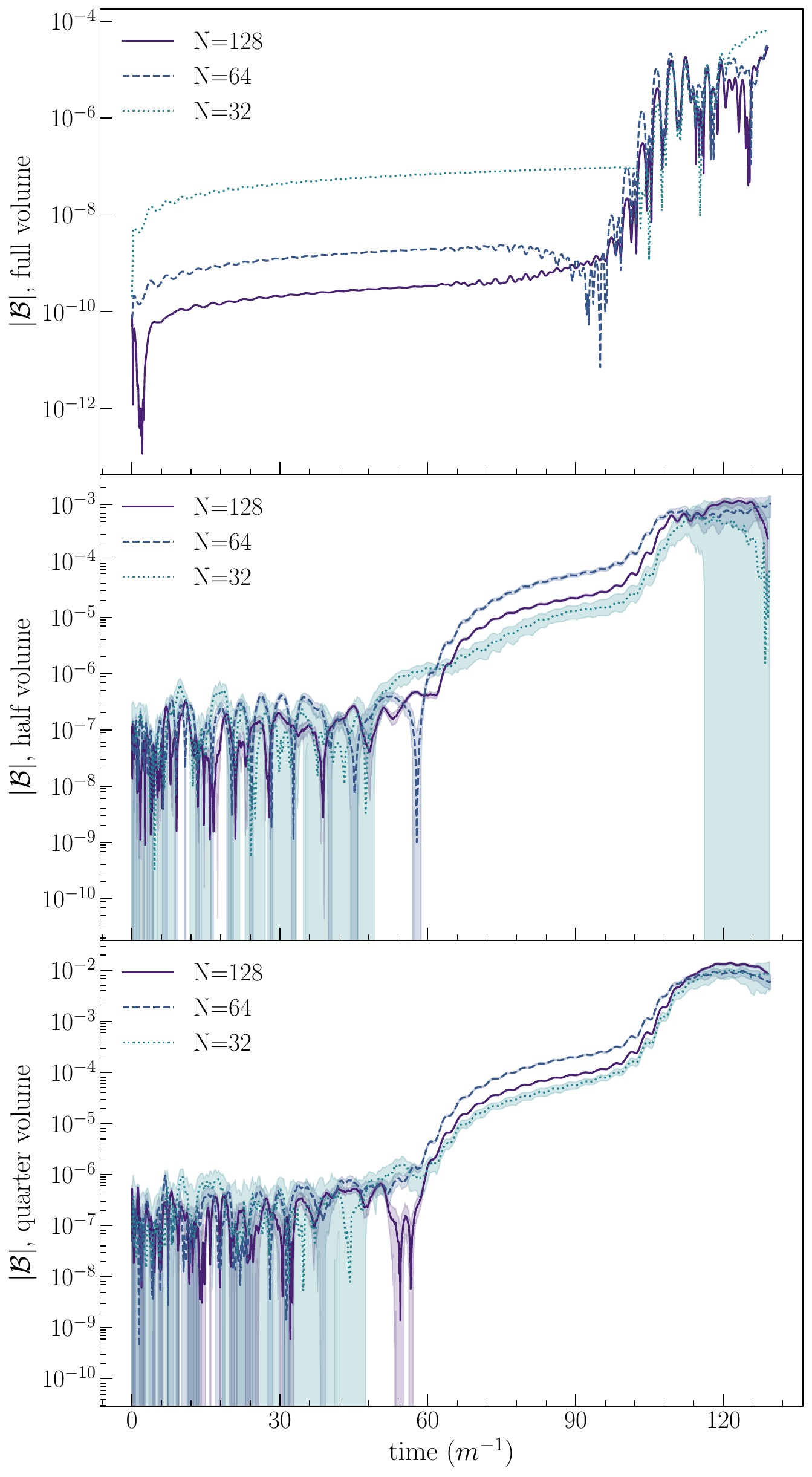}
    \caption{Convergence test of backreaction across resolutions $N=32$ (teal, dotted), $N=64$ (blue, dashed), and $N=128$ (purple, solid). The top panel shows the backreaction on the full volume for each of these resolutions, the middle panel shows the median backreaction on half-volume subdomains for each of these resolutions, and the lower panel shows the median backreaction on quarter-volume subdomains for each of these resolutions.}
    \label{fig:res-converge}
\end{figure}

In the upper panel of Fig.~\ref{fig:res-converge} we show $\mathcal{B}$ of the full volume for three different resolutions: $N=32$ (teal, dotted), $N=64$ (blue, dashed), and $N=128$ (purple, solid). This figure shows that the calculation of $|\mathcal{B}|$ reduces with an increase in resolution at early times $t\lesssim100$, while being of the same order of magnitude and equivalent for $N=64$ and $N=128$ at the end of the simulation. This implies that at early times, $|\mathcal{B}|$ as averaged over the whole box is likely dominated by numerical error and can be considered zero in a physical sense. Early times are dominated by the homogeneous mode, and since backreaction is zero at linear order, it is unsurprising that $\mathcal{B}$ is consistent with numerical error during this period. However, at late times of $t\gtrsim 100$, as power is transferred into high frequency modes and nonlinearity increases, the numerical error is subdominant to the physical violation of the Friedmann constraint, and the value of $\mathcal{B}$ has converged.

In middle and lower panels of Fig.~\ref{fig:res-converge} we show our calculations of the median $\mathcal{B}$ of half- and quarter-volume subdomains, respectively. Solid purple curves indicate the $N=128$ resolution, dashed blue curves indicate $N=64$, and dotted teal curves indicate $N=32$. For both the half- and quarter-volume subdomains, there is no significant shift in the median of $|\mathcal{B}|$ with a change in numerical resolution, and the differences are accounted for by the consideration that these are averages of a randomized subset rather than comparison between an exactly matching set. This acceptable degree of consistency indicates that our results for these subdomain sizes are converged.

\subsection{Box size}\label{appendix:box size}
\begin{figure}
    \centering
\includegraphics[width=0.95\linewidth]{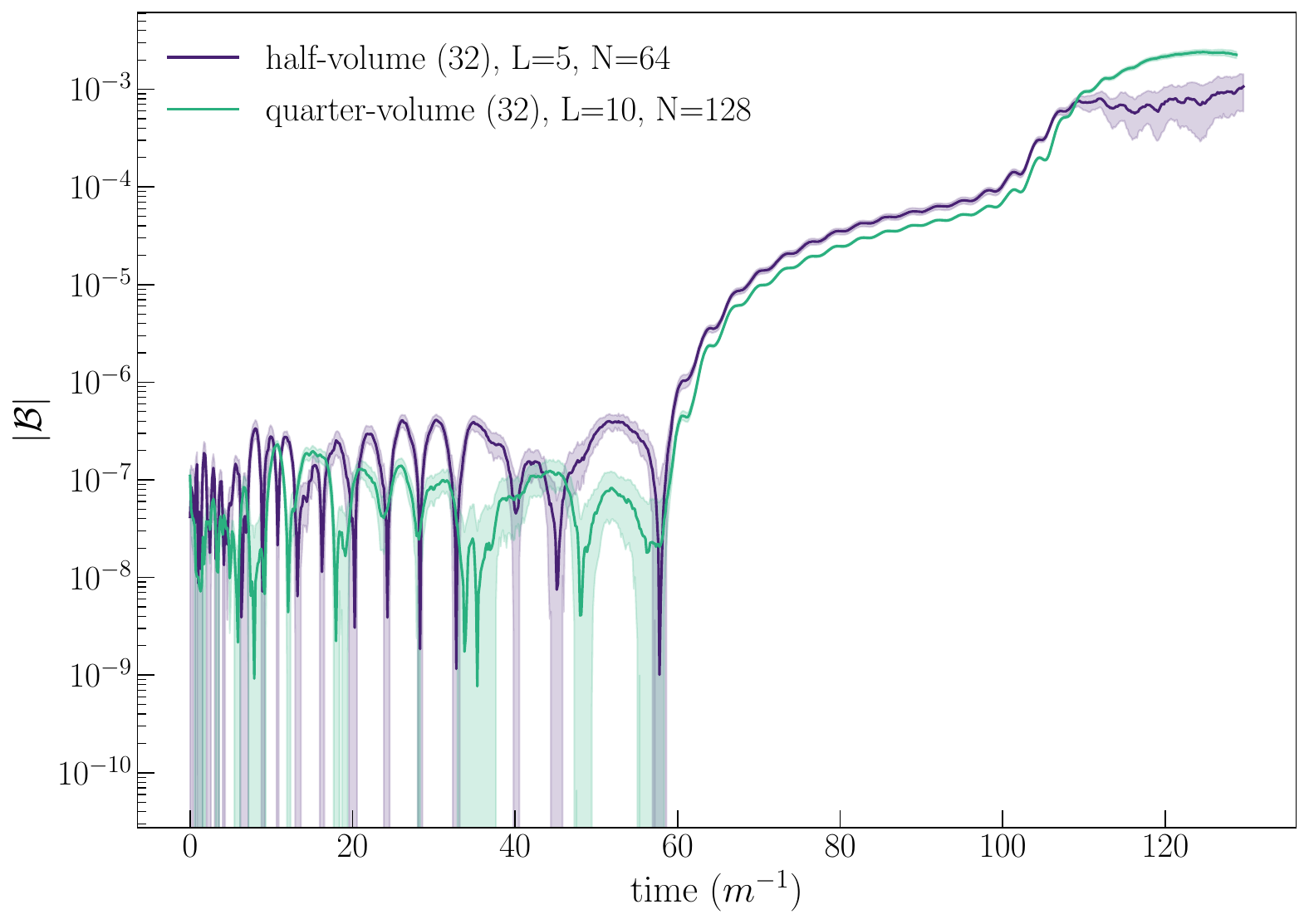}
    \caption{Comparison of median backreaction for different box sizes, as indicated in the legend. The purple line shows the average backreaction of half-volume of a $L=5$, $N=64$ simulation, while the teal line shows the average backreaction of quarter-volume from a $L=10$, $N=128$ simulation.}
    \label{fig:size-comp-half-quarter}
\end{figure}

We do not expect the choice of box size to greatly influence the results. A larger box allows for longer wavelength modes, which can contribute larger variations in gravitational potential~\cite{Giblin:2019nuv}. However, it is important to ensure that there are no issues arising from the setting of the simulation, such as boundary conditions. We use periodic boundary conditions, which should not influence the physics. 
Therefore, to check robustness of our results against box size, we compare the backreaction behavior in a box size of $L=10m^{-1}$ to the fiducial box size, $L=5m^{-1}$. We compare equivalent physical sized subdomains in two simulations, one with box size $L=10m^{-1}$ and $N=128$ and one with $L=5m^{-1}$ and $N=64$, with initial power spectrum cut-off of $\xi=1/32$ and $\xi=1/16$, respectively. We will compare the half-volume of the $L=5$ simulation and the quarter-volume of the $L=10$ simulation, such that both are subdomains of $32^3$ grid cells. As described in~\ref{ssec:subdom}, the number of subdomains is set by $N$, so there are 13,107 subdomains in the $N=64$, $L=5$ simulation and 104,857 in the $N=128$, $L=10$. 
In Fig.~\ref{fig:size-comp-half-quarter} we show this comparison, with the half-volume of $L=5,$ $N=64$ in purple and the quarter-volume of $L=10,$ $N=128$ in teal. To within a factor of two, the backreaction behavior is comparable throughout.

\subsection{Slicing condition}\label{appendix:slicing converge}
The choice of gauge in numerical relativity is arbitrary and should not impact any physically meaningful result. An important exception is in spatial averaging, since this inherently relies on the definition of spatial domains, which lie in space-time. Spatial averaging in simulations has been demonstrated to be explicitly dependent on the chosen gauge~\cite{Adamek2019,Giblin2019b}, so the identification of a `physically meaningful' gauge is important in this case. As discussed in the main text, we are interested in studying backreaction in traditional preheating simulations, which must choose some gauge. In this section we test the robustness of our results to the choice of foliation, using two additional similar slicings conditions. 
The fiducial slicing condition used for the results of the paper is given by Eqn.~\eqref{eq:main-slicing}. The other two slicing conditions we use are
\begin{equation}
\label{eq:slice1}
    \partial_t\alpha=-K/\alpha^2
\end{equation} and 
\begin{equation}
    \label{eq:slice2}
    \partial_t\alpha=1-\alpha.
\end{equation}
These slicing conditions were all chosen to keep $\alpha$ close to one in order to keep the simulation time coordinate close to cosmic time, which keeps the frequency of the homogeneous mode constant~\cite{Baumgarte:2010ndz,Baumgarte:2021skc}. Additionally, they maintain $\partial_t\alpha\sim 0$ such that our time steps remain mostly consistent through the simulation to maintain stability. 

Since changing the slicing inherently changes the time coordinate, to visualize different foliations together we need to transform the time coordinate. 
In BSSN, the line element is given by
\begin{equation}\label{appx:LE1}
    ds^2=-\alpha^2dt^2+\gamma_{ij}dx^idx^j,
\end{equation}
where each foliation corresponds to a different $\alpha(t)$ evolution. To match time coordinates, we transform each simulation to the following line element
\begin{equation}\label{appx:LE2}
    ds^2=-d\tau^2+\gamma_{ij}dx^idx^j,
\end{equation}
where the spatial coordinates $x^i$ are unchanged. Matching Eqns.~\eqref{appx:LE1} and~\eqref{appx:LE1}, and enforcing the initial times $t_{\rm ini}=\tau_{\rm ini}$, we arrive at
\begin{equation}
\label{eq:tau}
    \tau=\int\bar{\alpha}dt - dt.
\end{equation}
In practice, we perform this transformation using the coordinate time, $t$, and lapse, $\alpha$ from each simulation to scale the $x$-coordinate of our plots to $\tau$.

\begin{figure}
    \centering
    \includegraphics[width=0.95\linewidth]{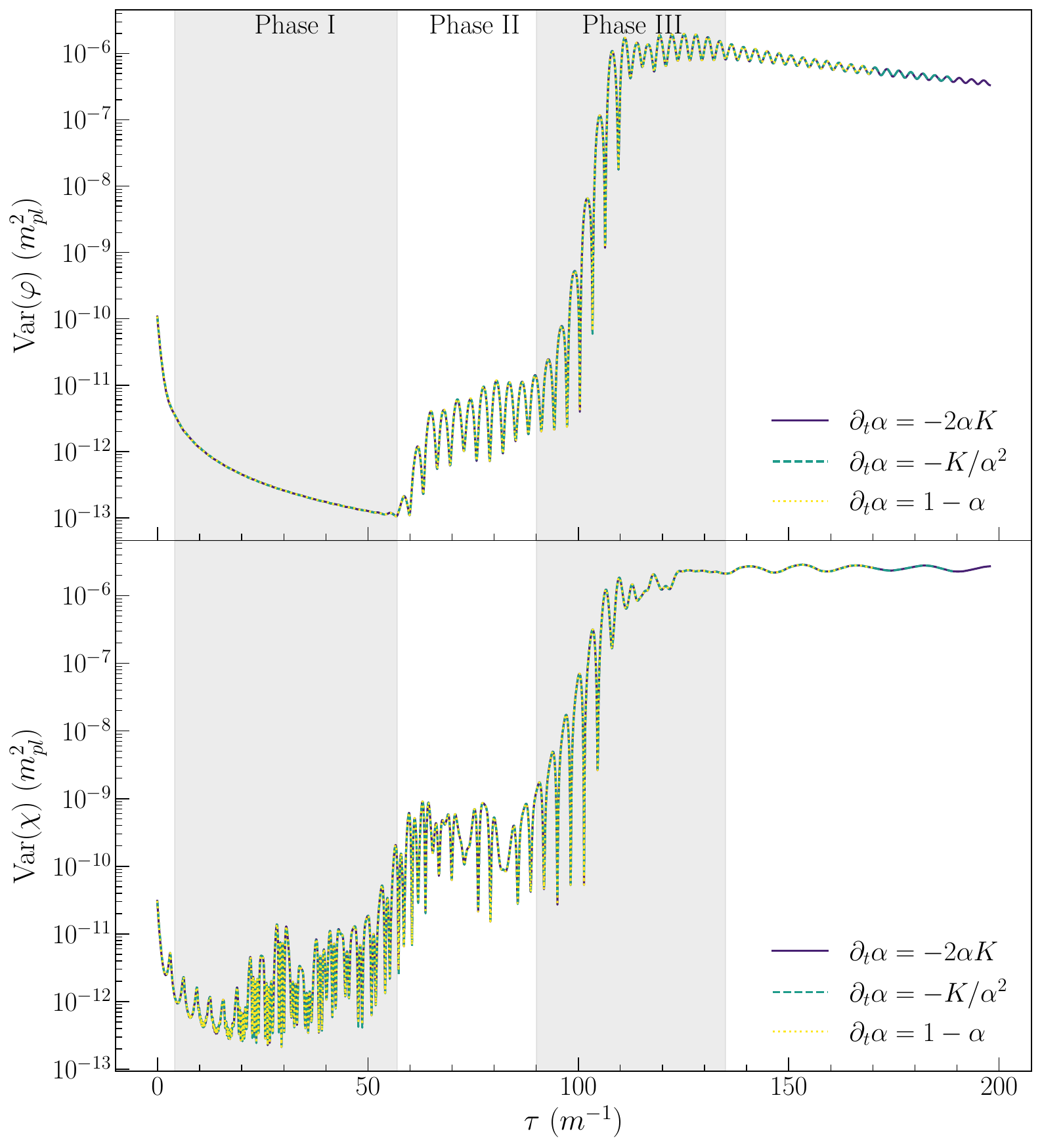}
    \caption{Comparison of the variance of the inflaton (upper panel) and of the coupled scalar field (lower panel) for three different foliations, as a function of scaled time coordinate $\tau$~\eqref{eq:tau}. The solid purple line shows the fiducial slicing condition~\eqref{eq:main-slicing}, while the dashed teal line shows the \eqref{eq:slice1} slicing condition and the dotted yellow line shows the \eqref{eq:slice2} slicing condition.}
    \label{fig:var-slicing}
\end{figure}

In Fig.~\ref{fig:var-slicing}, we show the variance of the inflaton (upper panel) and coupled scalar field (lower panel), as a function of the scaled time coordinate, $\tau$. The solid purple line shows the fiducial slicing condition~\eqref{eq:main-slicing}, while the dashed teal line shows the \eqref{eq:slice1} slicing condition and the dotted yellow line shows the \eqref{eq:slice2} slicing condition. The variance of the inflaton and coupled field match across all three different foliations, as expected. 

\begin{figure}
    \centering
    \includegraphics[width=0.95\linewidth]{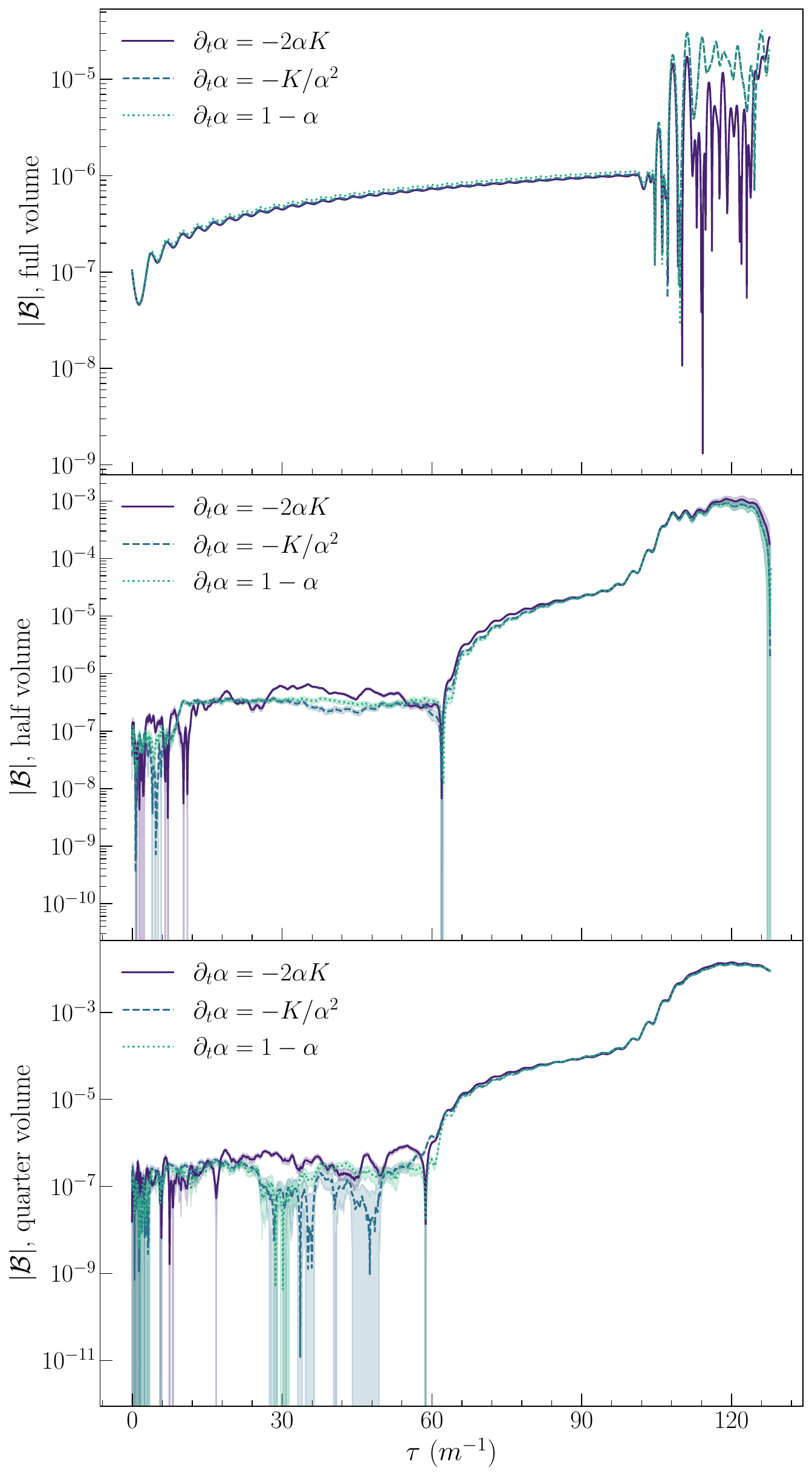}
    \caption{Comparison of backreaction for three different foliations, plotted as a function of scaled time coordinate $\tau$, scaled by $\eqref{eq:tau}$. In all panels, the fiducial slicing condition~\eqref{eq:main-slicing} is given by the solid purple line, the~\eqref{eq:slice1} slicing condition by the dashed blue line, and the~\eqref{eq:slice2} slicing condition by the dotted teal line. The top panel shows backreaction across the full volume, the middle panel shows the median backreaction across half-volume subdomains, and the lower panel shows the median backreaction across quarter-volume subdomains.}
    \label{fig:slice-converge}
\end{figure}

Next, we compare $\mathcal{B}$, given by Eqn.~\eqref{eq:Bdef}, for each of the three choices of slicing condition. In Fig.~\ref{fig:slice-converge}, the top panel shows the backreaction for the full volume, the middle panel shows the median backreaction of half-volume subdomains, and the lower panel shows the median backreaction of quarter-volume subdomains. In all three panels, the fiducial slicing condition~\eqref{eq:main-slicing} is given by the solid purple line, the~\eqref{eq:slice1} slicing condition by the dashed blue line, and the~\eqref{eq:slice2} slicing condition by the dotted teal line. Backreaction across the full volume matches within the same order of magnitude for all foliations by the end of preheating. The median backreaction across half- and quarter-volumes matches near-identically at late times ($\tau\gtrsim80$). This indicates that we can trust our results are robust to the choice of foliation.

\section{Constraint violation}\label{appendix:constraint}
The Hamiltonian constraint \eqref{eq:ham} is satisfied identically for any exact solution of Einstein's equations. For numerical solutions, any violation of this constraint can thus be attributed to approximations in initial data or numerical error accumulated during the simulation. Thus, it is common to use the constraints as a measure of numerical error to track how close we are to a true solution of Einstein's equations. If the violation of the Hamiltonian constraint becomes `unsatisfactory' (discussed below), we can no longer trust the physics of the simulation, even if the code remains stable and does not crash.

The violation of the Hamiltonian constraint, Eqn.~\eqref{eq:ham}, is 
\begin{equation}
    H\equiv\mathcal{R} + K^2 - K_{ij}K^{ij} - \frac{16\pi\rho}{m_\text{pl}^2},
\end{equation}
which GABERel calculates at each grid point in the code. It then takes the Euclidean mean (i.e., the average given by Eqn.~\eqref{eq:avg_def} assuming $\gamma=1$ everywhere) at each time step. We compare the level of violation for changes in resolution as well as the different foliations.

\subsection{Resolution}\label{appendix:res constraint}
Here, we check to ensure the violation in the Hamiltonian constraint (which is expected to be dominated by numerical finite-difference error from the simulation evolution) reduces as we increase resolution. 

\begin{figure}[H]
    \centering
    \includegraphics[width=0.95\linewidth]{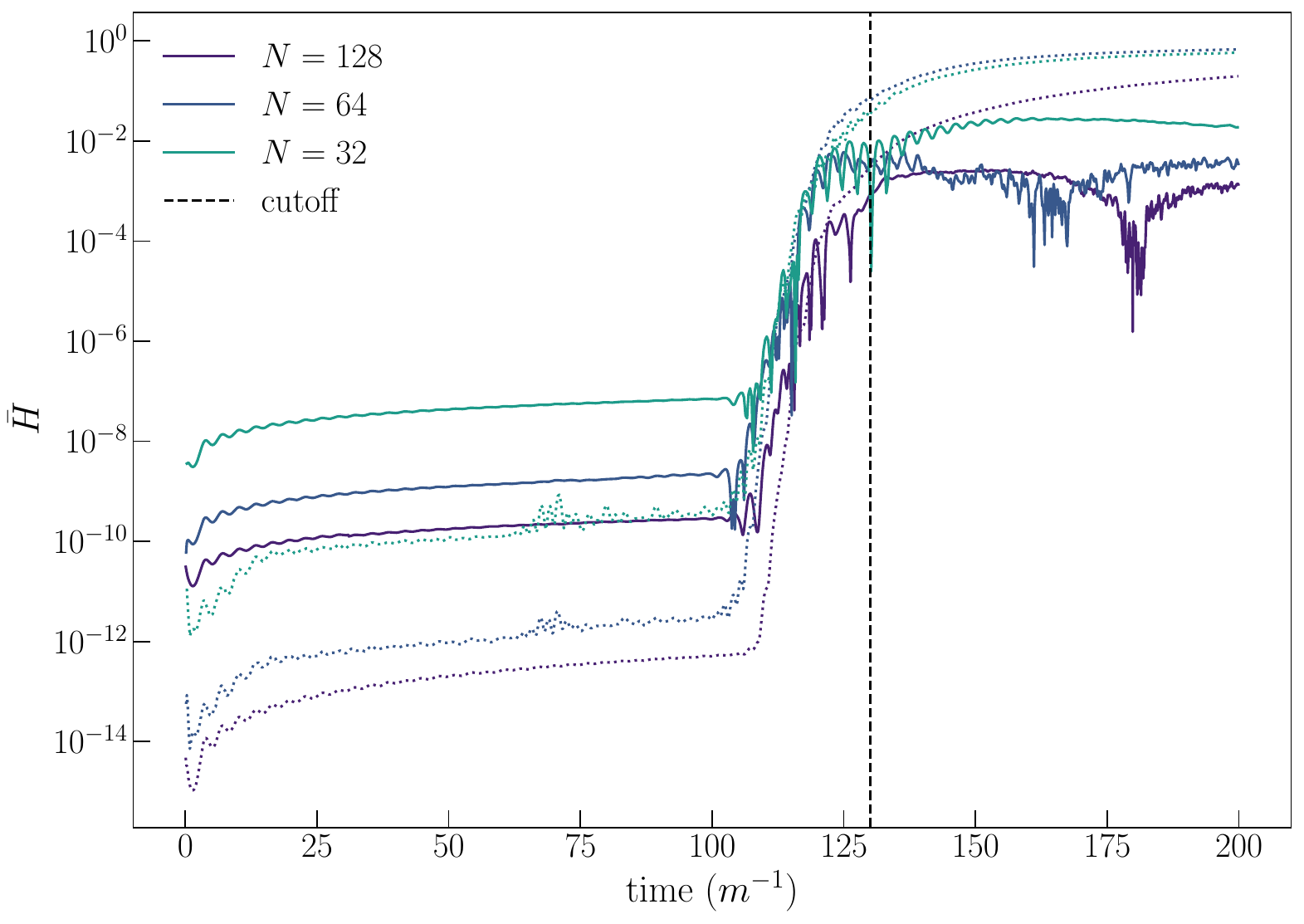}
    \caption{The Hamiltonian constraint for three different resolutions. Each color corresponds to a resolution, with the solid curve showing the mean Hamiltonian constraint and the dotted curve showing its variance. The fiducial simulation, with a resolution of $N=128$, is shown by the purple curves. The blue curves are for $N=64$, and the teal curves are for $N=32$. The variance surpasses $1\%$ at $t\approx130$ (dashed black vertical line), at which point we become wary of the simulation results.}
    \label{fig:ham-constraint-res}
\end{figure}

In Fig.~\ref{fig:ham-constraint-res} we show the mean Hamiltonian constraint over time for three different resolutions, with the fiducial resolution $N=128$ in purple, $N=64$ in blue, and $N=32$ in teal. We see it converges toward zero as we increase resolution, and it remains both bounded and satisfied to below 1\% for the entirety of the simulation. 
Dotted curves in Fig.~\ref{fig:ham-constraint-res} of the same color as solid curves show the variance, as defined by Eqn.~\ref{eq:variance}, in the constraint violation for the respective resolution. Both the mean and the variance of the violation increase rapidly after $t\sim 100$ when power is transferred to the inhomogeneous modes of the scalar field. This is to be expected since as smaller-scale modes develop, they are sampled by fewer grid cells and thus their evolution will contain a larger error. While both the mean and the variance remain bounded through the end of the simulation at $t\sim 200$, the variance surpasses our chosen 1\% violation limit at $t\sim 130$. Thus, results past this point must be taken with a grain of salt, and we choose to err on the side of caution by cutting off the results shown in this work at this point.

\subsection{Slicing condition}\label{appendix:slicing constraint}
\begin{figure}
    \centering
    \includegraphics[width=0.95\linewidth]{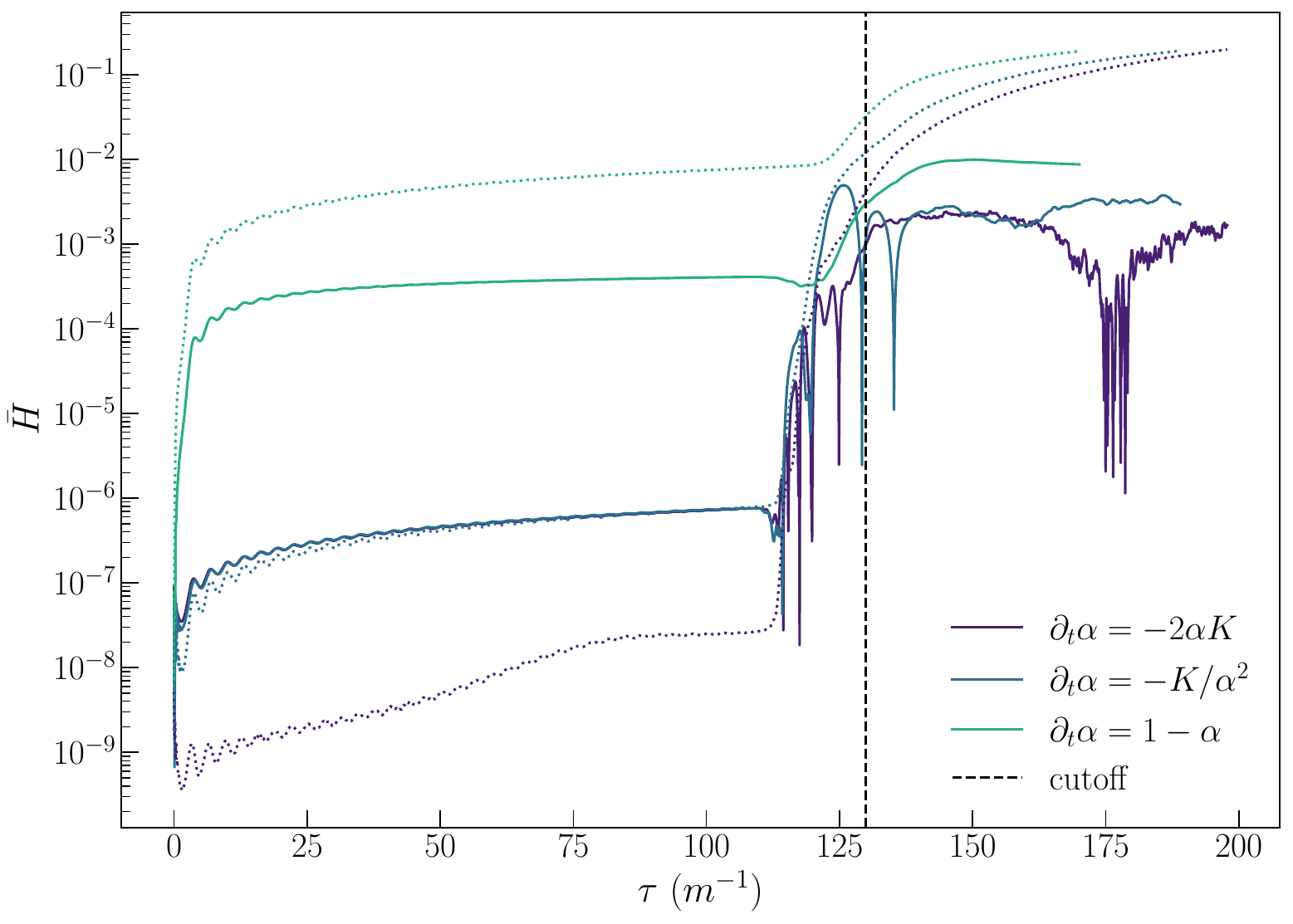}
    \caption{The Hamiltonian constraint for the full runtime, for three different foliations. Solid curves show the mean violation, while the dotted curves show its variance. The fiducial simulation, with slicing condition given by Eqn.~\eqref{eq:main-slicing} is shown by the purple curves. The blue curves use the slicing condtion given by Eqn.~\eqref{eq:slice1}, while the teal curves use the slicing condition given by Eqn.~\eqref{eq:slice2}. While the latter starts at a higher point, all three show similar behavior of where they are bounded.}
    \label{fig:ham-constraint-slicing}
\end{figure}

Fig.~\ref{fig:ham-constraint-slicing} shows the mean Hamiltonian constraint violation for the three different choices of foliation we examined: the fiducial slicing condition~\eqref{eq:main-slicing} (purple curve), along with~\eqref{eq:slice1} (blue curve), andn~\eqref{eq:slice2} (teal curve). While the slicing condition\eqref{eq:slice2} is initially higher, all three remain bounded and under $1\%$ at the start. The dashed curves in Fig.~\ref{fig:ham-constraint-slicing} show the variance of the Hamiltonian constraint, indicating that by $t\sim130$ (dashed black vertical line) the constraint has surpassed the $1\%$ limit we have conservatively set.

\bibliography{refs}

\end{document}